\documentclass{article}
\usepackage{graphicx}
\usepackage{amsthm}
\usepackage{amsfonts}

\newcommand{\head}[1]{\centerline{\LARGE\bf #1}}
\newcommand{\ignore}[1]{}
\newtheorem{problem}{Open problem}

\begin{document}

\head{Recent advances in open billiards}
\head{with some open problems}
\vskip .25cm
\centerline{Carl P. Dettmann, Department of Mathematics, University of Bristol, UK}
\begin{abstract}
Much recent interest has focused on ``open'' dynamical systems, in which a classical map or flow is
considered only until the trajectory reaches a ``hole'', at which the dynamics is no longer considered.
Here we consider questions pertaining to the survival probability as a function of time, given an initial measure on phase space.  We focus on the case of billiard dynamics, namely that of a
point particle moving with constant velocity except for mirror-like reflections at the boundary,
and give a number of recent results, physical applications and open problems.
\end{abstract}

\section{Introduction}
A mathematical billiard is a dynamical system in which a point particle moves with
constant speed in a straight line in a compact
domain ${\cal D}\subset \mathbb{R}^d$ with a piecewise smooth%
\footnote{The amount of smoothness required depends on the context; an early result of Lazutkin~\cite{L73} required 553 continuous derivatives, but recent theorems for chaotic billiards
typically require three continuous derivatives except at a small set of singular points~\cite{BG}
while polygonal billiards and most explicit examples are piecewise analytic.  For an attempt in
a (non-smooth) fractal direction, see~\cite{LN}.}
boundary $\partial {\cal D}$ and making mirror-like%
\footnote{That is, the angle of incidence is equal to the angle of reflection.  Popular alternative
reflection laws include that of outer billiards, also called dual billiards, in which the
dynamics is external to a convex domain, see for example~\cite{DF09}, and Andreev billiards used
to model superconductors~\cite{CHK09}; we do not consider these here.}
reflections whenever it reaches the boundary.  We can assume that the speed and mass are both equal
to unity.  In some cases it is convenient to use periodic boundary conditions with obstacles
in the interior, so $\mathbb{R}^d$ is replaced by the torus $\mathbb{T}^d$.  Here we will
mostly consider the planar case $d=2$.
Billiards are
of interest in mathematics as examples of many different dynamical behaviours (see Fig.~1 below)
and in  physics as models in statistical mechanics 
and the classical limit of waves moving in homogeneous cavities; more details for both
mathematics and physics are given below.  An effort has been made to keep the
discussion as non-technical and self-contained as possible; for further definitions and
discussion, please see~\cite{CM,DY,G98}.  It should also be noted that the references contain
only a personal and very incomplete selection of the huge literature relevant to open billiards,
and that further open problems may be found in the recent reviews~\cite{Bun08,DY,N08,S10,S08}

\begin{figure}
\vspace{-1cm}
\centering
\includegraphics[width=100pt]{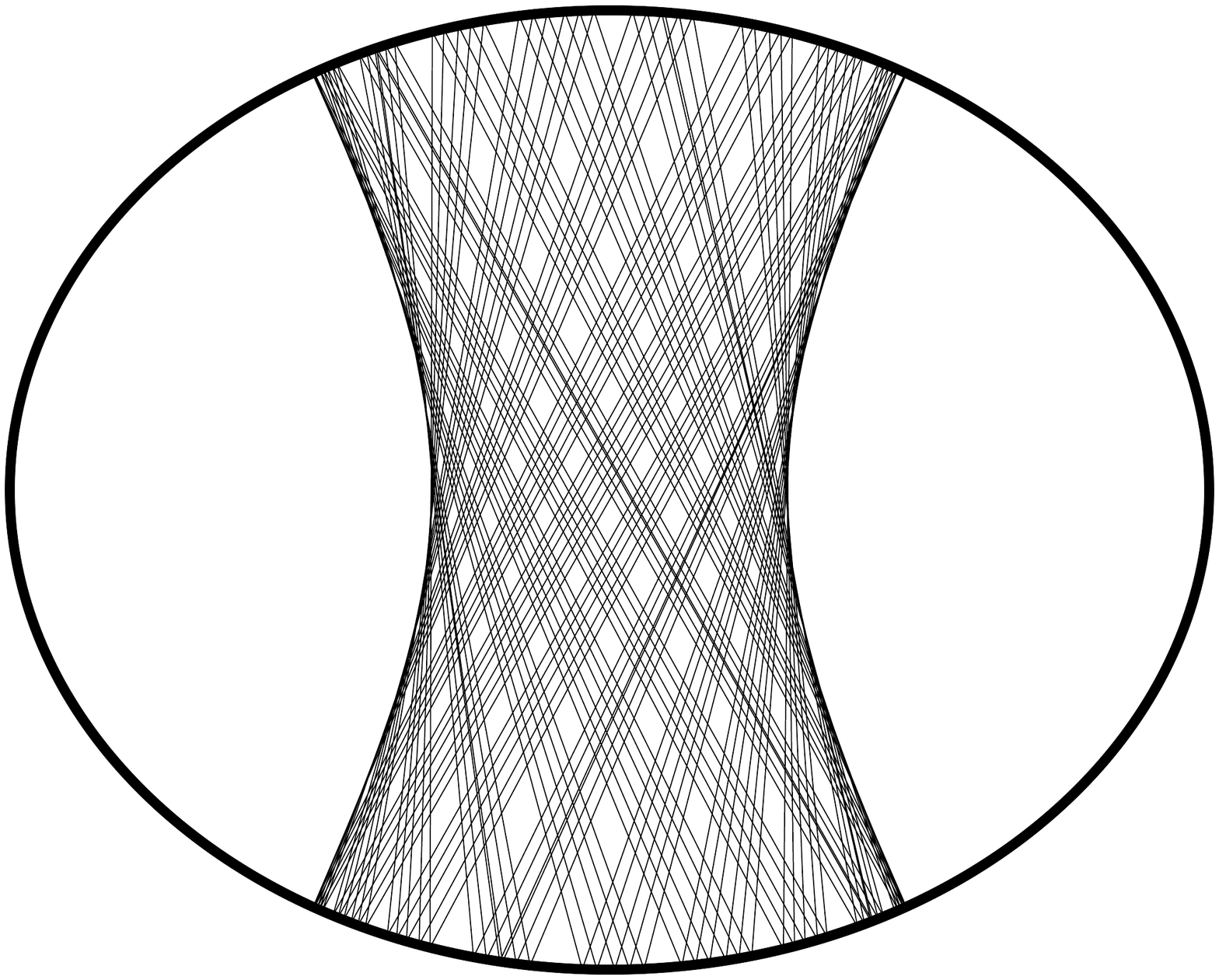}
\includegraphics[width=100pt]{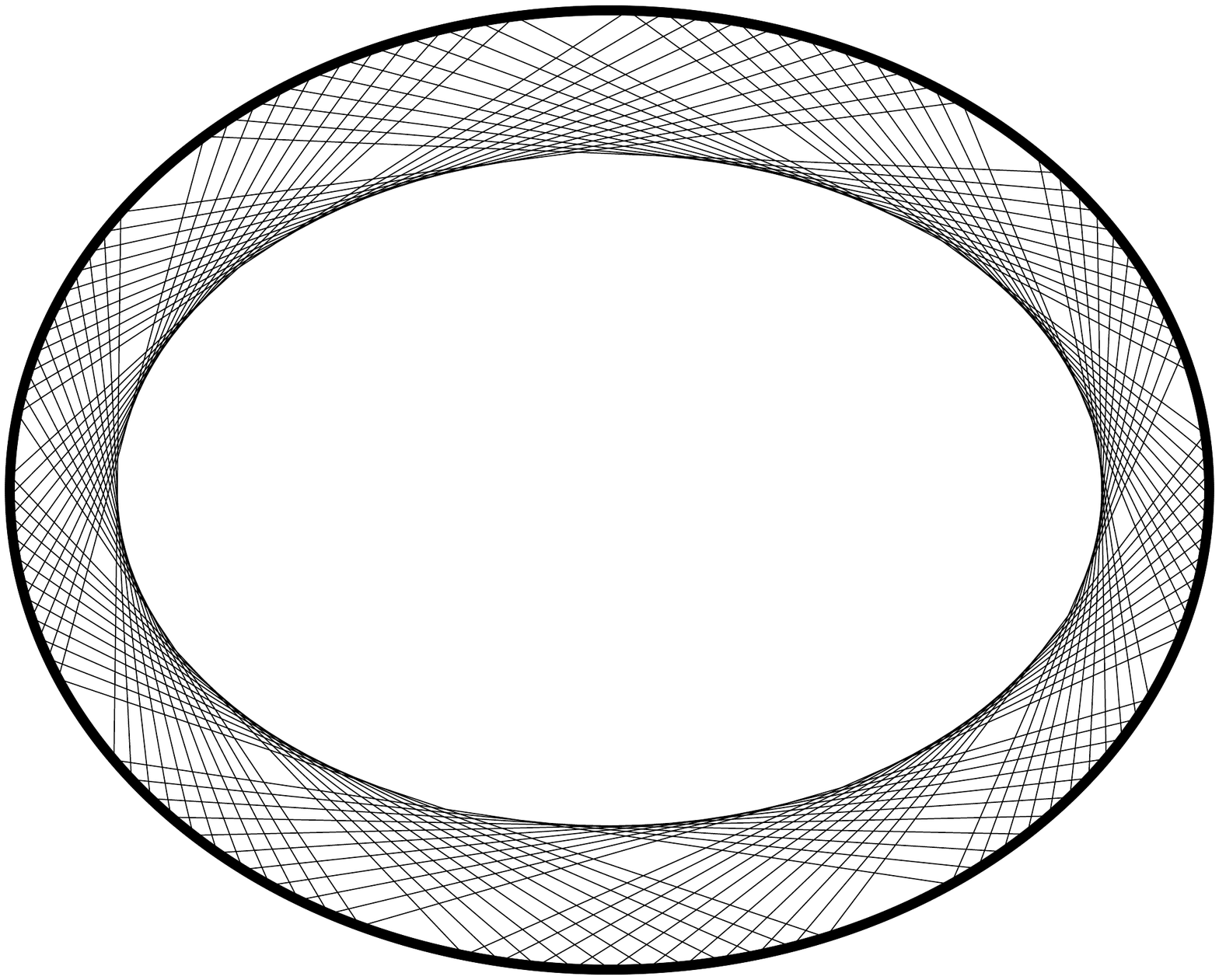}
\includegraphics[width=100pt]{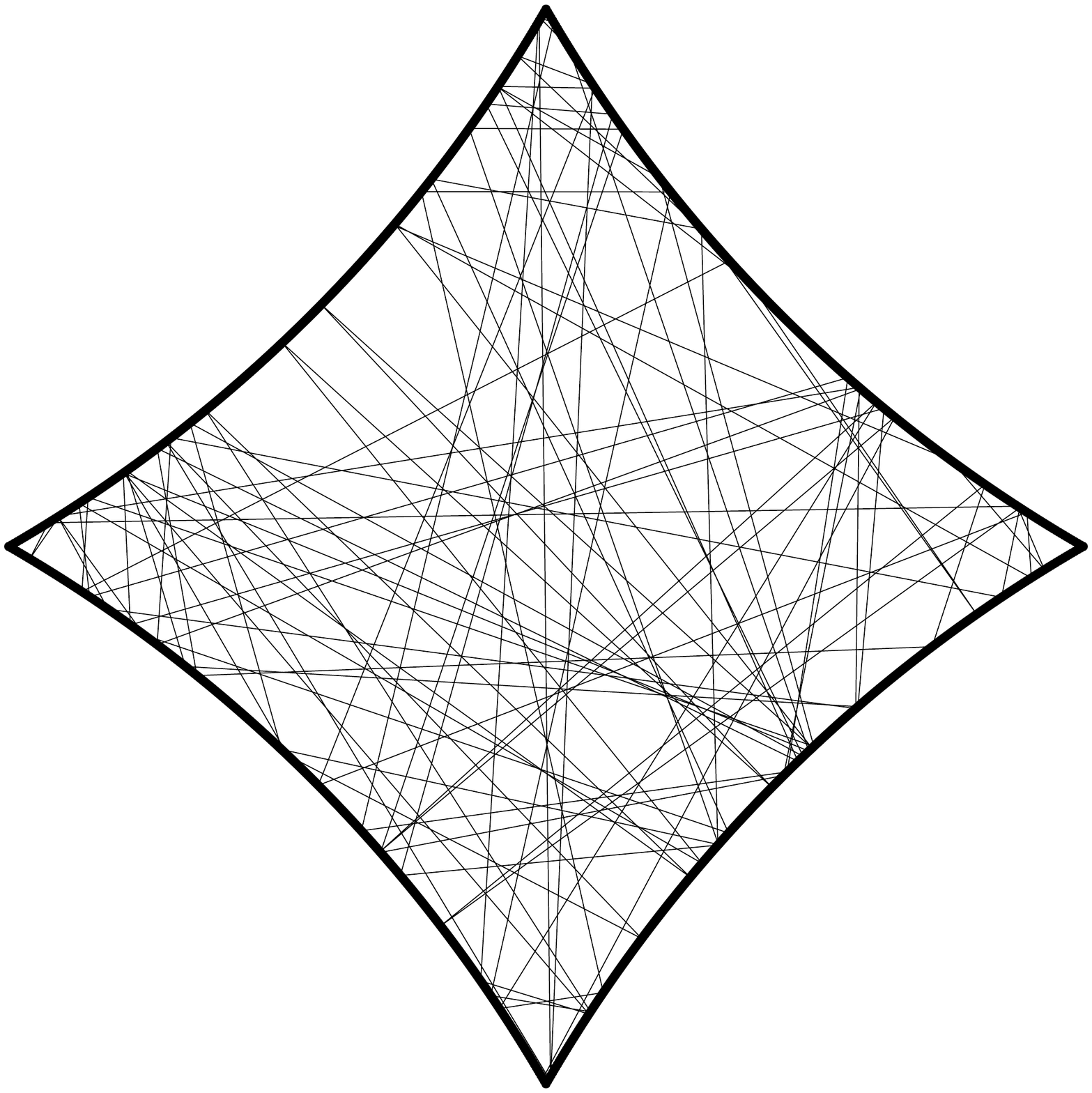}\\
\vspace{-1.5cm}
\includegraphics[width=100pt]{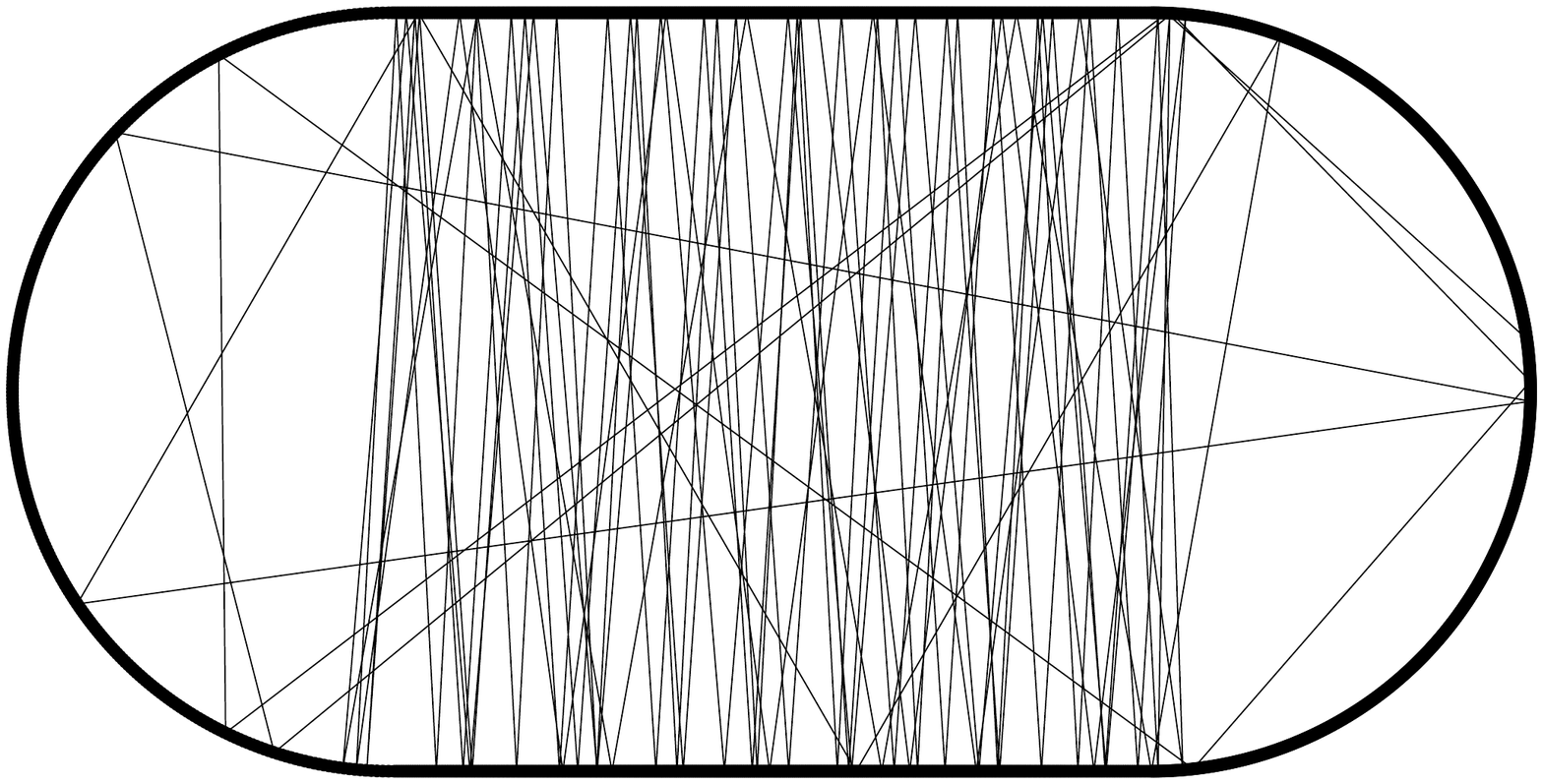}
\includegraphics[width=100pt]{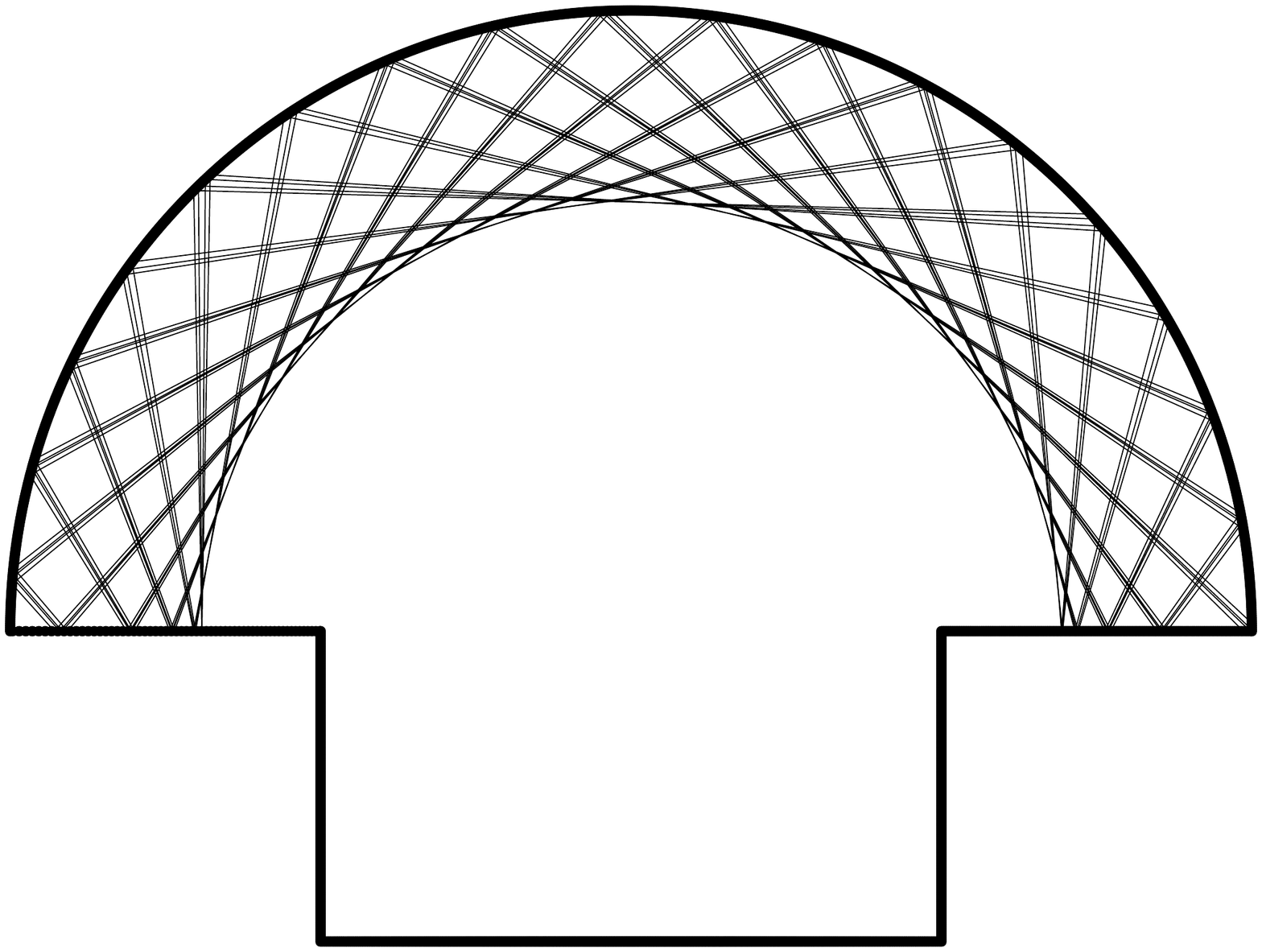}
\includegraphics[width=100pt]{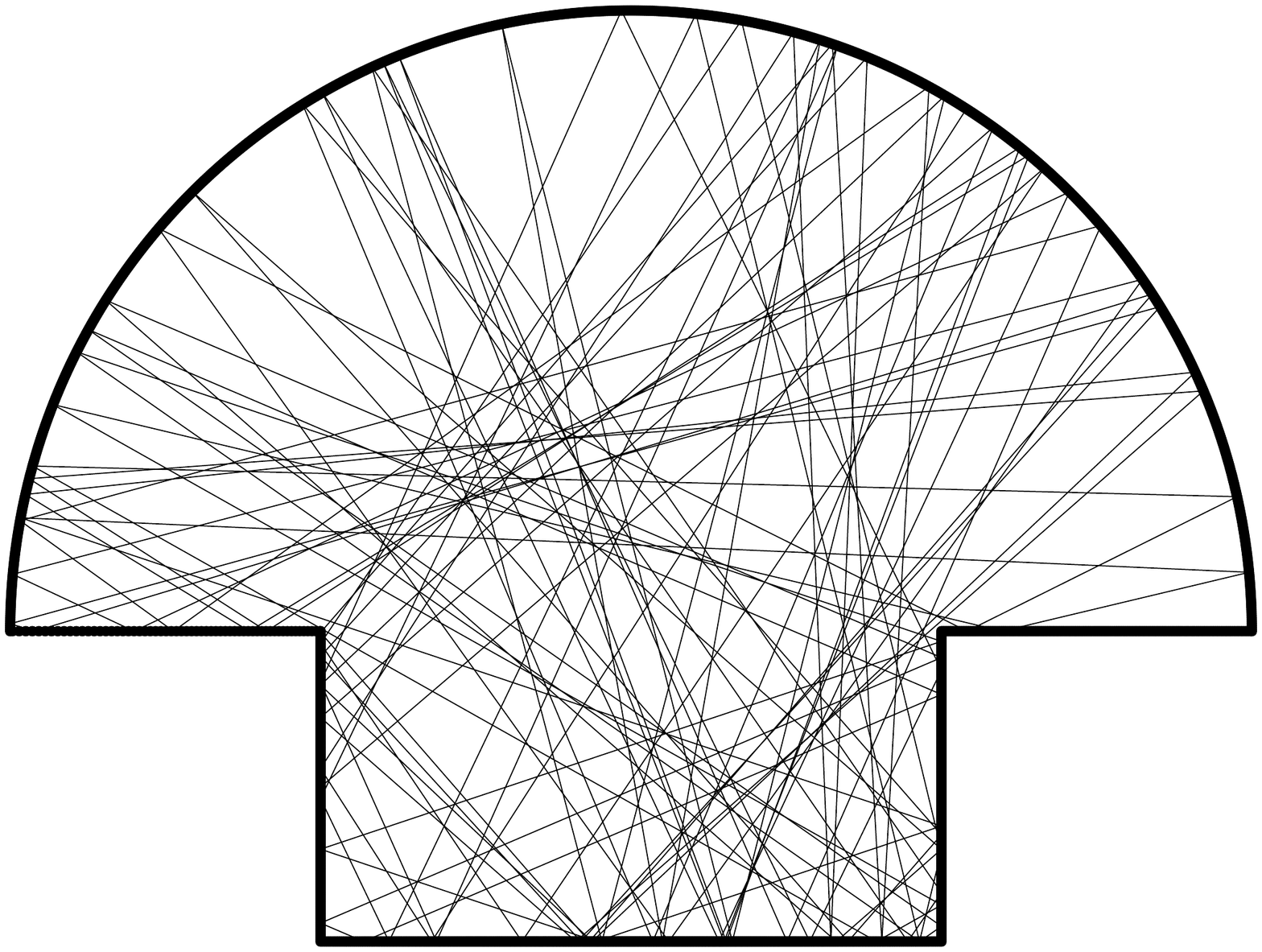}
\vspace{-1cm}
\caption{Geometry vs. dynamics.  An integrable elliptical billiard (top left and centre) has two
types of regular motion depending on the initial condition, a dispersing diamond (top right) is
chaotic, a defocusing stadium (bottom left) is intermittent, with orbits switching between
regular and chaotic motion, and a mushroom (bottom centre and right) is mixed,
with regular or chaotic motion depending on the initial condition.  Apart from
the ellipse, all these are constructed from circular arcs and straight lines; many other
examples with more subtle dynamical distinctions exist.  See Sec.~\protect\ref{s:dyn} for
discussion of all these cases.{\vspace{-0.3cm}}}
\end{figure}

The structure of this article is as follows.  In Secs.~2 and 3 we consider general work on
closed and open dynamical systems respectively. In Sec.~4 we consider each of the main classes
of billiard dynamics, and in Sec.~5 we consider physical applications.  Sec.~6 returns to a
more general discussion and outlook. 

\section{Closed dynamical systems}\label{s:gen}
In this section we introduce some notation for mathematical billiards, as well as informal
descriptions of properties used to characterise chaos in billiards and more general systems,
with pointers to more precise formulations in the literature.  Much relevant work on open
systems applies to chaotic maps different from billiards, so this notation is designed to
be applicable both to billiards and to more general dynamical systems.  A readable introduction
to the subject of (closed) billiards is given in~\cite{T} with more details of chaotic
billiards in~\cite{CM}.  

Below, $|\cdot |$ will denote the size of a set, using the Lebesgue measure of the
appropriate dimension.%
\footnote{All sets are assumed to be measurable with respect to the relevant measure(s).}
We denote the dynamics by $\Phi^t:\Omega\to\Omega$ for either flows
(including the case of billiards) or maps.  For flows we have $t\in\mathbb{R}$ while for maps
we have $t\in\mathbb{Z}$. $\Phi^t$ also naturally acts on subsets $A\subseteq\Omega$.  For
non-invertible maps and $t<0$, action on points is undefined, but action on sets gives the
relevant pre-image. The phase space $\Omega$ for the billiard flow consists of all%
\footnote{\label{f:inf} Billiard dynamics cannot usually be continued uniquely if the particle
reaches a corner, so strictly speaking we need to exclude a zero measure set of points that do this
at some time in the past and/or future.  Another barrier to defining $\Phi^t$ is where
there are infinitely many collisions in finite time;~\cite{CM} states conditions under which
this is impossible, roughly that for $d=2$ there are finitely many corners, and focusing
(convex) parts of the boundary are sufficiently ($C^3$) smooth, have non-zero curvature, and
do not end in a cusp (zero angle corner). This is the case for almost all billiards considered
in the past; exceptions may be interesting to consider in the future.}
particle positions $q\in {\cal D}$ and momentum directions $p\in \mathbb{S}^{d-1}$ except that
we need an extra condition to make $\Phi^t$ single-valued at collisions: For definiteness assume
that $p$ points toward the interior of $\cal D$ if $q$ is at the boundary.

For many calculations it is more convenient to put a Poincar\'e section on the billiard boundary
(or other convenient hypersurface for general flows),
thus considering the ``Birkhoff'' map $F^n:M\to M$ describing
$n\in\mathbb{Z}$ collisions, and acting on a reduced phase space $M$ consisting of points
$x\in\partial {\cal D}$ and inward pointing $p$.  The map and flow are related by the
``roof function'', the time $\tau:M\to\mathbb{R}^+$ to the next collision, so that $F(x)=\Phi^{\tau(x)}(x)$ for $x\in M$.   

We now mention a few%
\footnote{The properties mentioned are those commonly discussed in the literature of open
dynamical systems including billiards, and tend to be measure theoretic rather than topological.}
properties used to describe the chaoticity of dynamical systems; more
details including subtle differences between flows and maps can be found in~\cite{CM}.
A hyperbolic map is one where the Lyapunov exponents (exponential expansion and contraction
rates of infinitesimal perturbations) $\lambda_i$ are all non-zero almost everywhere, and in the
flow case there is a single zero exponent associated with the flow direction. For a uniformly
hyperbolic system the relevant statement is true everywhere, with uniform bounds on the
exponents and associated constants.  For smooth uniformly hyperbolic (Anosov)
systems the dynamics is controlled
by unstable and stable manifolds in phase space corresponding to the positive and negative
Lyapunov exponents respectively, yielding Sinai-Ruelle-Bowen (SRB) invariant measures smooth in
the unstable directions, and dividing phase space into convenient entities called
Markov partitions. Hyperbolicity that is nonuniform and/or nonsmooth has similar properties,
but requires more general and detailed techniques~\cite{BP}.

For ergodic properties we choose an invariant measure%
\footnote{That is, $\mu(\Phi^{-t}A)=\mu(A)$ for any $A\subseteq\Omega$ and any $t$ (positive
for noninvertible maps).} $\mu$. For billiards there are natural
``equilibrium'' invariant probability measures $\mu_{\Omega}$ and $\mu_M$.
$\mu_{\Omega}$ is given by the usual (Lebesgue) measure on the phase
space $\Omega$, and $\mu_M$ is given by the product of measure on the
boundary and the components of momentum parallel to the boundary.
For example, in two dimensions $d\mu_M=ds\; dp_\parallel/(2|\partial{\cal D}|)
=\cos\psi\;ds\; d\psi/(2|\partial{\cal D}|)$,
where $s$ is arc length, $p_\parallel$ is momentum parallel to the boundary and $\psi$
is angle of incidence, that is, the angle between the momentum following a collision
and the inward normal to the boundary.
For integrable billiards such as the ellipse of Fig.~1, this is one of many
smooth invariant measures defined by arbitrary smooth functions on the level sets of
the conserved quantity, while for ergodic billiards such as the stadium or diamond of
Fig.~1, it is the only smooth invariant measure.  

Given an invariant measure $\mu$, recurrence is the statement that $\mu$-almost all
trajectories return arbitrarily close to their initial point; it is guaranteed by the
Poincar\'e recurrence theorem if $\mu$ is finite as above. Ergodicity is the statement that
all invariant sets have zero or full measure, which implies that time averages of an integrable
phase function are for almost all (with respect to $\mu$) initial conditions given by its phase
space average over $\mu$.  Mixing is the statement that the $\mu$-probability of visiting region
$A$ at time zero and $B$ at time $t$ are statistically independent in the limit $|t|\to\infty$
for sets $A,B\subset\Omega$.  There is an equivalent statement in terms of decay of correlation
functions such as found in Eq.~(\ref{e:corr}).  Stronger ergodic properties are Kolmogorov mixing
(K-mixing) and Bernoulli.  We have Bernoulli $ \Rightarrow$ K-mixing $ \Rightarrow$ mixing $ \Rightarrow$
ergodicity $ \Rightarrow$ recurrence.

Further statistical properties build on the above. These include rates of decay of correlation
functions for mixing systems and moderately regular (typically H\"older continuous) phase
functions and central limit theorems for time averages of phase functions.
See~\cite{BM,CZ} for a discussion of recent results in this direction.

Finally, the Kolmogorov-Sinai (KS) entropy $h_{KS}$ is a measure of the unpredictability of the
system. For a closed, sufficiently smooth map with a smooth invariant measure it is equal to
the sum of the positive Lyapunov exponents (if any); this is called Pesin's formula, and has been
proved for some systems with singularities including billiards.  All K-mixing systems have $h_{KS}>0$.  

Note that many of these descriptors of chaos are logically independent.  In the case of
billiards, open problem~\ref{o:mixing} conjectures that polygonal billiards may be mixing,
but they are certainly not hyperbolic, having zero Lyapunov exponents everywhere.
The recently proposed ``track billiards'' \cite{BD09} are hyperbolic but not ergodic or mixing.
The stadium of Fig.~1 and the section below on defocusing billiards is Bernoulli~\cite{CM} but
has slow decay of correlations \cite{BM} and a non-standard central limit theorem \cite{BG06}.

\section{Open dynamical systems}
Open dynamical systems are reviewed in both the mathematical \cite{DY} and physical \cite{AT}
literature.  Most of the mathematical studies of open dynamical systems have considered strongly
chaotic systems, such as piecewise expanding maps of the interval or Anosov (uniformly hyperbolic)
maps in higher dimensions.  Note that there are a variety of conventions used in the literature
to describe open systems.

An open dynamical system contains a ``hole'' $H\subset \Omega$ at which the particle is absorbed and
no longer considered.  $H$ may have more than one connected component (several ``holes''), and
may be on the boundary (ie a subset of $M$) or in the interior, but should be piecewise smooth and of
dimension one less than $\Omega$ for flows, and of the same dimension as $\Omega$ for maps.
Here we allow a particle to be injected at the hole, so it is absorbed only when it reaches the hole
at strictly positive time.  If we denote by $\Omega_t$ the subset of $\Omega$ that does not reach
the hole by time $t$,
\begin{equation}
\Omega_t=\{x\in \Omega : \Phi^s(x)\notin H,\quad \forall s\in(0,t]\}
\end{equation}
a typical question to ask is that given a set of initial conditions distributed according to
some probability measure $\mu_0$ at time $t=0$, what is the probability $P(t)=\mu_0(\Omega_t)$
that the particle survives until time $t$?  How does this probability behave as a function of $t$,
the initial measure $\mu_0$, the hole location $H$, its size relative to the billiard boundary $h=|H|/|M|$ (for general maps, $|H|/|\Omega|$), and the shape of the billiard ${\cal D}$?

In mathematical literature \cite{LM,PS} the term ``open
billiard'' has been used to incorporate additional conditions. In terms of our notation the
outer boundary of ${\cal D}$ is a strictly convex set forming the ``hole'' through which particles
escape; ${\cal D}$ also excludes three or more strictly convex connected obstacles in its interior
satisfying a non-eclipsing condition, that is, the convex hull of the union of any
two obstacles does not intersect any other obstacles. This ensures that there is a trapped
set of orbits that never escape, hyperbolic and with a Markov partition, leading to a relatively
good understanding \cite{CERRS,G98,LM,PS,S90}. Billiards satisfying these conditions will be
denoted here as ``non-eclipsing'' billiards.

For strongly chaotic systems (including the Anosov maps mentioned above and dispersing billiards
with finite horizon discussed below) we expect that $P(t)$ decays with time as described by
an (exponential) escape rate
\begin{equation}\label{e:gamma}
\gamma=-\lim_{t\to\infty}\frac{1}{t}\ln P(t)
\end{equation}
which exists and is independent of a reasonable class of initial measures $\mu_0$.%
\footnote{In \cite{DY} some examples are given where existence is shown for $\mu_0$ equivalent to
Lebesgue with density bounded away from 0 and $\infty$; this condition is probably too restrictive.}
We recall that $t$ corresponds either to discrete or continuous time depending on the context, with
(the more physical) continuous time implied for billiards except where otherwise stated.

We now discuss conditionally invariant measures; see also \cite{DY}.
The renormalised evolution $\Phi^t_H$ of the initial measure $\mu_0$ is defined by its
action on sets $A\subseteq\Omega$:
\begin{equation}
\mu_t(A)=(\Phi_H^t\mu_0)(A)=\frac{\mu_0(\Phi^{-t}(A)\cap\Omega_t)}{\mu_0(\Omega_t)}
\end{equation}
It is easy to check that $\Phi_H^t$ satisfies the semigroup
property $\Phi_H^s\circ\Phi_H^t=\Phi_H^{s+t}$ for non-negative $s$ and $t$.  If $\mu_t$ approaches
a limit $\mu_\infty$ at which $\Phi_H^t$ is continuous (in a suitable topology), this limit
is conditionally invariant, ie $\Phi_H^t\mu_\infty=\mu_\infty$ for all $t\geq 0$,
and gives the escape rate
\begin{equation}
\mu_\infty(\Omega_t)=e^{-\gamma t}
\end{equation}
also for all $t\geq 0$.

\cite{KL01} makes the important point that for open systems, conditionally
invariant measures of $\Phi$ (projected onto $M$) and $F$, while supported on the same set,
are generally not equivalent measures, even when the roof function $\tau$ is smooth, in constrast
to the closed case.  In other words escape properties of the flow do not follow trivially from
those of the map.  For billiard calculations, it is usually best to work first with the collision map $F$ (projecting any interior holes onto the collision space $M$),
then try to incoporate effects of the flow.  For incorporation of flow effects in proofs of
statistical properties of closed billiards, see \cite{BM,CM} and in calculation of escape rates and
averages of open systems, see \cite{BD07,G98,L}.  

Discussion in \cite{DY} demonstrates that conditionally invariant measures can have properties
similar to the SRB measures of closed hyperbolic systems, for example being smooth
in the unstable direction.  For non-invertible maps the measures satisfying the property of
conditional invariance can however be highly non-unique.

The conditionally invariant measure $\mu_\infty$ is supported on the set of points with infinite
past avoiding the hole, however most of these points reach the hole in the future.  There is an
invariant set of points never reaching the hole in past or future called the repeller; the
natural invariant measure on this set is defined by the limit
\begin{equation}
\nu(A)=\lim_{t\to\infty} e^{\gamma t}\mu_{\infty}(A\cap\Omega_t)
\end{equation}
Pesin's formula generalises in the open case to the ``escape rate formula''~\cite{DY,G98}
\begin{equation}\label{e:erf}
\gamma=\sum_{i:\lambda_i>0}\lambda_i-h_{KS}
\end{equation}
where the quantities on the right hand side are defined with respect to $\nu$.%
\footnote{Proofs of the escape rate formula typically replace the sum of
Lyapunov exponents by the (more general) Jacobian of the dynamics restricted to unstable directions.
For two dimensional billiards, at most one Lyapunov exponent is positive.}
The escape rate formula has very recently
been shown for a system that is not uniformly hyperbolic~\cite{BDM}, however there is wide
scope for further development both in maps without uniform hyperbolicity and in flows.

\begin{problem}
How general is the escape rate formula, Eq.~(\ref{e:erf})?
\end{problem}

There has been a very recent explosion of interest in the mathematics of open dynamical
systems \cite{AB,BY,DWY,KL,PS}, again mostly restricted to
strongly hyperbolic maps as above.  In \cite{BY,KL} it is shown that $\gamma/h$ can reach a limit,
the local escape rate, as the hole shrinks to a point,
and that the limit depends on whether the point is periodic.
For example consider the map $f(x)= 2x\;({\rm mod}\;1)$ and a point $z$ of minimal period $p$
so that $f^p(z)=z$, together with a sequence of holes $H_n$ of size $h_n$ and each containing $z$,
with corresponding escape rates $\gamma_n$.  Then there is a local escape rate
\begin{equation}\label{e:ler}
\lambda_z=\lim_{n\to\infty}\frac{\gamma_n}{h_n}=1-2^{-p}
\end{equation}  
If the point $z$ is aperiodic, then $\lambda_z$ is equal to $1$.  For more general 1D maps, the
$2^{-p}$ is replaced by the inverse of the stability factor $|(d/dx) f^p(x)|$ at $x=z$ and the
invariant measure of the map (uniform in the above example) needs to be taken into account.

This and related results are, however available only for piecewise expanding maps and closely
related systems, although ``it is conceivable that in the near future this result could be
applied e.g. to billiards.'' \cite{KL}.  The very recent
work \cite{DWY} on the periodic Lorentz gas, a dispersing billiard model (see below)
shows that $\gamma$ (defined with respect to collisions, not continuous time)
is well defined for sufficiently small holes, corresponding
to a limiting conditionally invariant measure that is independent of the initial measure for
a relatively large class of the latter, also that $\gamma\to 0$ as $h\to 0$, but does not say
anything about $\gamma/h$.  This leads to the open problem:

\begin{problem}\label{o:loc}
Local escape rate: Proof of a formula similar to Eq. (\ref{e:ler}) for sufficiently chaotic
billiard models.
\end{problem}

Note that the holes typically considered in billiards, consisting of a set small in position
but with arbitrary momentum,%
\footnote{Note that other possibilities occur in applications (Sec. \ref{s:appl}):
Escape for trajectories
with sufficiently small angle of incidence but at any boundary point is relevant to microlasers,
and escape with probability depending on the boundary material is relevant to room acoustics.}  
include phase space distant from any short periodic orbit contained in the
hole.  Following the discussion in \cite{BD07}, the effects of the periodic orbit may then
appear at a higher power of the hole size, compared to the above piecewise expanding map, so
for example the formula might look something like $\gamma=h+h^2\gamma_2+O(h^3)$ with
$\gamma_2$ depending on properties of the shortest periodic orbit contained in the hole.
Also note (see open problem \ref{o:scaling} below) that if there is a local escape rate, $P$ behaves
like $e^{-\lambda_z ht}$, in other words like a function of the combination $ht$ in this limit.

Another important question posed in this work is that of optimisation: how to choose a hole
position to minimise or maximise escape, for example $\gamma$ or even the whole
function $P(t)$ \cite{AB,BY}.

\begin{problem}
Optimisation: Specify where to place a hole to maximise or minimise a suitably defined measure of
escape.
\end{problem}

In the above papers, slow escape is related to placing the hole on a short periodic orbit or on
the part of phase space with the lowest ``network load.''  The main question is how these criteria
generalise to open dynamical systems (including billiards) with distortion, nonuniform hyperbolicity
or no hyperbolicity.  

As will become clear below, the
existence of an exponential escape rate $\gamma$ and nontrivial conditionally invariant
measure $\mu_\infty$ is only expected in fairly special cases: there are many systems
including billiards with slower (or occasionally faster) decay of the survival probability.
For some systems with slower eventual decay of $P(t)$, a cross-over from approximately
exponential decay at short times to algebraic decay at late times has been observed numerically \cite{AT}.  We now turn to our main focus, that of open billiards, returning to general
open dynamical systems in the applications section.

\section{Open billiards}\label{s:dyn}
Now we consider open billiards, categorised according to dynamical properties of the
corresponding closed case.  The initial measure $\mu_0$ for an open billiard is normally
given by the equilibrium measure for the flow or Poincar\'e map, $\mu_\Omega$ or $\mu_M$
respectively.  An alternative is to consider injection at the hole by a restriction of the
equilibrium measure, in other words transport through the billiard; in many cases this has
little effect on the long time properties \cite{AT}.

A few papers discussed here and in the statistical mechanics section below
consider billiards on infinite domains, usually consisting of
an infinite collection of non-overlapping obstacles.  If both the billiard and
any hole(s) are periodic, this is equivalent to motion on a torus; an example is \cite{DWY}.
However, if the billiard is aperiodic, or the hole is not repeated periodically, the above
definitions break down because the equilibrium billiard measure is not normalisable.
The above definition of ergodicity makes sense using an infinite invariant measure,
but other properties including mixing and survival probability are problematic.  Also,
the Poincar\'e recurrence theorem fails and so recurrence properties need to be proved.
One approach to defining $P(t)$ in the case of a non-repeated hole might be to choose an
initial measure $\mu_0$ with support in a bounded region, somewhat analogous to the finite (but
rather arbitrary) outer boundary of the domain ${\cal D}$ chosen for non-eclipsing
billiards above.

\paragraph{Integrable billiards}
A few billiards are integrable, ie perfectly regular, namely
the ellipse, rectangle, equilateral triangle and related cases \cite{T}.  In the integrable
circle with a hole of angle $2\pi h$ in the boundary \cite{BD05}, the leading order coefficient
of $P(t)$ was obtained exactly; in particular the statement
\begin{equation}
\lim_{h\to 0}\lim_{t\to\infty}h^{\delta-1/2}(tP(t)-\frac{1}{\pi h})=0
\end{equation}
for all $\delta>0$ is equivalent to the Riemann Hypothesis, the greatest unsolved problem
in number theory.%
\footnote{\label{f:RH} The Riemann hypothesis is the statement that all the complex zeros of
the Riemann zeta function, the analytic continuation of $\zeta(s)=\sum_{n=1}^\infty n^{-s}$ lie
on the line $Re(s)=1/2$, and is related to the distribution of prime numbers \cite{C03}.}
If instead both limits are taken simultaneously so that $ht$ is constant, the survival
probability numerically appears to reach a limiting function.

\begin{problem}\label{o:scaling}
Scaling: For what billiards and in what limits does $P(t)$ reduce to
a function of $ht$ for a family of billiards with variable hole size $h$?
\end{problem}

\paragraph{Polygonal billiards}
The survey \cite{G96} gives a good introduction to polygonal billiards.  A billiard collision
involving a straight piece of boundary is equivalent to free motion in an ``unfolded'' billiard
reflected across the boundary as in a mirror, using multiple Riemann sheets if the reflection
leads to an overlap.
General polygonal billiards with angles rational multiples of $\pi$ can be unfolded into
flat manifolds with nontrivial topology and conical singularities (from the corners),
called translation surfaces, which are currently an active area of interest \cite{AF08}.  In such
billiards each trajectory explores only a finite number of directions, however with this
restriction is typically ergodic but not mixing \cite{U}.  There has been recent progress on
showing recurrence for infinite rational polygonal billiards~\cite{T10}
 
Very little is known about the case of irrational angles \cite{JBR,V}, except that
usual characteristics of chaoticity such as positive Lyapunov exponents are not possible here.
We have the well known open question \cite{S09}:

\begin{problem}\label{o:tri}
Existence of periodic orbits: Do all triangular (or more generally polygonal) billiards
contain at least one periodic orbit?
\end{problem}

Note that this admits arbitrary directions; it is easy to construct rational polygonal
billiards for which a particular directional flow is never periodic. Also, there are a number
of cases including rational and acute triangles for which periodic orbits can be explicitly
constructed.  Periodic orbits in polyhedra are considered in \cite{Bed08}.

Another open problem has the interesting feature that mathematical literature conjectures
that it is false \cite{G96}, while physical literature conjectures that it is true \cite{CP};
recent work is found in \cite{U}:

\begin{problem}\label{o:mixing}
Mixing property: Is it possible for a polygonal billiard (necessarily with at least one
irrational angle) to be mixing?
\end{problem}

There is very little work on open polygonal billiards given the paucity of knowledge about the
closed case.  In \cite{DC} it was found numerically that $P(t)$ decayed as $C/t$ in an irrational
polygonal billiard where at least one periodic orbit was present, and vanished at finite time
in a directional flow for a rational polygonal billiard containing no periodic orbits.  So we have

\begin{problem}
Is the asymptotic survival probability $P(t)$ as $t\to\infty$ in polygonal billiards
entirely determined by the neighbourhoods of non-escaping periodic orbits?
\end{problem}

\paragraph{Dispersing billiards}
Curved boundaries of billiards lead to focusing (convergence) or dispersion (divergence)
of an initially parallel set of incoming trajectories, depending on the sign of the curvature.%
\footnote{
Zero curvature, ie inflection points and similar, can be problematic; see footnote \ref{f:inf}.
The main exception is where part of the boundary is exactly flat.  A class of billiards
allowing some inflection points that has been studied in detail is that of semi-dispersing
billiards \cite{CS,S08}.}
\footnote{
In three or more dimensions the curvature at a boundary point generally depends on the plane
of the incoming trajectory; this phenomenon is known as astigmatism and causes difficulties in
the theory, hence many results are weaker and/or require additional assumptions; see
for example \cite{BT,PS}.}
We have already noted that one focusing billiard, the ellipse, is integrable; more generally all
sufficiently smooth strictly convex billiards are not ergodic \cite{L73}.  However there are
a number of other classes of billiards which are not only ergodic, but have much stronger
chaotic properties, to which we turn; \cite{CM} provides a good introduction and~\cite{S10,S08}
recent reviews.

Beginning with the Sinai billiard \cite{S70}
which used a convex obstacle in a torus to ensure that all collisions were dispersing,
there have been many new constructions of chaotic billiards and techniques to prove stronger
statistical properties of existing billiards.  For smooth dispersing cases such as Sinai's
billiard, ergodic properties now include the Bernoulli property.
The theory is technical due to singularities in the dynamics, requiring additional techniques
for more difficult classes of singularity:  All dispersing billiards have trajectories tangent
to one or more parts of the boundary; perturbed trajectories either miss that part of the
boundary entirely, or collide with an outgoing perturbation proportional to the square root
of the incoming one, that is, infinite linear instability.  In addition,
the boundary may contain corners, in most cases leading to discontinuities in the dynamics.
In the case of cusps (zero angle corners) the number of collisions near the cusp
in a finite time is unbounded.  A further complication occurs in motion on a torus if some
trajectories never collide
(``infinite horizon condition''),%
\footnote{In the context of infinite billiards, there has recently been proposed a ``locally finite
horizon condition'', in which all straight lines pass through infinitely many scatterers, but
the free path length is unbounded, see~\cite{T10}.  Here it is also assumed that the minimum
distance between obstacles be bounded away from zero.}
due to an accumulation of singularity sets near such orbits.
Proof of exponential decay of correlations of the billiard map was possible using Young
tower constructions of the late 1990s which generalise the concept of a Markov partition,
but exponential decay of correlations for the flow is still conjectured;
recent results for advanced statistical properties and treatment of continuous time may be
found in \cite{CZ,BM} and some open problems in \cite{S08}.

An example of a dispersing billiard with corners is the diamond shown in Fig. 1.
This was the numerical example in \cite{BD07}, where the exponential escape rate $\gamma$ as
above can be expanded in powers of the hole size $h$ using series of correlation functions, as
can the differential escape rate of one vs two holes.  For example, the expression for the
latter, following a derivation that is far from rigorous but consistent with numerical tests, is
\begin{equation}\label{e:corr}
\gamma_{AB}=\gamma_A+\gamma_B-\frac{1}{\langle \tau\rangle}
\sum_{j=-\infty}^{\infty}\langle (u_A) (F^j\circ u_B)\rangle+\ldots
\end{equation}
where $A$ and $B$ label the holes, $\tau$ is the time from one collision to the next,
$u$ is a zero mean phase function (that is, $\langle u\rangle=0$)
equal to $-1$ on the relevant hole and $h\tau/\langle \tau\rangle$
elsewhere, angle brackets indicate integration with respect to normalised equilibrium measure
on the billiard boundary $\mu_M$, and $F$ is the billiard map.
The omitted terms have third or higher order correlations, and are expected to be of order
$h^3$ as long as $A$ and $B$ do not contain points from the same short periodic orbit.
Convergence of the sum over $j$ follows from sufficiently fast decay of correlations,
which do not need to be exponential.  An extended diamond geometry was used as a model for
heat conduction in \cite{GG}.

\begin{problem}
Give a rigorous formulation of the escape rate of dispersing billiards with small holes
along the lines of Eq. (\ref{e:corr}).
\end{problem}

Dispersing billiards, which include the non-eclipsing billiards defined above, also provide a
likely setting for solving the open problems in Sec. \ref{s:gen}.

\paragraph{Defocusing and intermittent billiards}
In addition to the dispersing billiards, strongly chaotic billiards may be constructed with
the defocusing mechanism of Bunimovich, in which trajectories that leave a focusing piece
of boundary have a sufficient time to defocus (ie disperse) before reaching another curved
part of the boundary; for a recent discussion see \cite{BG}.  In this case there are
difficulties due to ``whispering gallery'' orbits close to a focusing boundary with finite but
unbounded collisions in a finite time.  Some concepts and results can be derived from the
Pesin theory of nonuniformly hyperbolic dynamics \cite{BP,CM}.

An example of a defocusing
billiard is the stadium of Fig. 1; note that while it is ergodic and hyperbolic,
the presence of ``bouncing ball'' orbits between the straight
segments leads to intermittent quasi-regular behaviour, leading to some weaker statistical
properties, so for example while the system is mixing, decay of correlations is now $C/t$ for
both the map and flow, leading to a non-standard central limit theorem with $\sqrt{n\ln n}$
(rather than $\sqrt{n}$) normalisation \cite{BG06,BM}.

The open stadium has been considered in \cite{AHO,DG}; this is a good model for intermittency
due to ``bouncing ball'' motions between the straight segments.  The first paper discusses
scaling behaviour associated with evolution of measures at long times and small angles close
to the bouncing ball orbits, while the second finds an explicit expression for the leading
coefficient of the survival probability
\begin{equation}
\lim_{t\to\infty}tP(t)=\frac{(3\ln 3+4)((a+h_1)^2+(a-h_2)^2)}{4(4a+2\pi r)}
\end{equation}
for a stadium with horizontal straight segments $x\in(-a,a)$ containing a small hole for
$x\in(h_1,h_2)$ with $-a<h_1<h_2<a$, and semicircles of radius $r$.  Note that this
approaches a constant as $h\sim h_2-h_1\to 0$, thus demonstrating an example where fixed
$ht$ is {\em not} a correct scaling limit (compare open problem \ref{o:scaling}).

The stadium is the most famous defocusing billiard, but many of its properties are due to
the intermittency arising from the bouncing ball orbits, or more generally a family of marginally
unstable periodic orbits, rather than the defocusing mechanism per se; defocusing billiards need
not have such orbits \cite{BG}.  One source of interest and also difficulty with the stadium is
the fact that orbits that leave the bouncing ball region are immediately reinjected back,
with an angle that is approximately described using an independent stochastic process \cite{AHO}.
Other reinjection mechanisms are possible, depending on the properties of the curvature of
the boundary approaching the end points of the bouncing ball orbits.  In some cases small
changes of the boundary of a stadium lead to a breakdown of ergodicity~\cite{Gthesis}, which
of course will also affect relevant escape problems.

\begin{problem}
Give a comprehensive characterisation of the dynamical (including escape) properties of
marginally unstable periodic orbits in stadium-like billiards in terms of their reinjection
dynamics.
\end{problem}

\paragraph{Mixed billiards}
Typical billiards are expected to have mixed phase space, that is, chaotic or
regular depending on the initial condition, in a fractal hierarchy according to
Kolmogorov-Arnold-Moser (KAM) theory \cite{Br}.
Much remains to be understood about this generic case; it has been conjectured~\cite{AT,CK}
that in the case of area preserving maps the long time survival probability associated with
stickiness near the elliptic islands decays as $t^{-\alpha}$ with a universal $\alpha\approx 2.57$.

\begin{problem}
Is there a universal decay rate in generic open billiards?
\end{problem}

One fruitful
line of enquiry has been the introduction of mushroom billiards by Bunimovich; these
have mixed phase space, but with a smooth boundary between the regular and chaotic regions.
See Fig. 1 and \cite{Bun08}.  The chaotic region of most of these
is intermittent (``sticky'') due to embedded marginally unstable periodic
orbits \cite{ACWM}, however an example of a non-sticky mushroom-like billiard was
given in \cite{Bun08}. The decay of the survival probability in the case that the hole
is in the chaotic region may be significantly slowed both by stickiness due to marginally
unstable periodic orbits, and by the boundary itself.%
\footnote{Recent unpublished result of the author in collaboration with O. Georgiou.}

\begin{problem}
Characterise the escape properties of the boundary between regular and chaotic regions
in mushroom and similar billiards.
\end{problem}

\section{Physical applications}\label{s:appl}
\paragraph{Statistical mechanics}
One important application of billiards is that of atomic and molecular interactions, for which steep
repulsive potential energy functions can be approximated by the hard collisions of billiards.
A system of many particles undergoing hard collisions corresponds to a high dimensional billiard.
This type of billiard is described as semi-dispersing since during a collision of two particles
the other particles are not affected, so there are many directions in the high dimensional collision
space with zero curvature; however there has been recent progress in demonstrating ergodicity
(conjectured by Boltzmann); see \cite{CS,S08}.  Note, however, that ergodicity may be broken by
arbitrarily steep potential energies approximating the billiard \cite{RR}.

\begin{problem}
For physical systems with many particles which are predominantly chaotic, how prevalent and
important are the regular regions (``elliptic islands'') in phase space? \cite{Br,Bun08,RR}.
\end{problem}

Systems of many hard particles have also been instrumental in the discovery of a fascinating
connection between microscopic and macroscopic dynamical effects, that of Lyapunov modes,
although later found in systems with soft potentials; for a review see \cite{YR}.  There are
many other connections between dynamics in general and statistical mechanics \cite{D,MPRV}.

Low-dimensional billiards imitating hard-ball motion are popular for understanding statistical
mechanics, particularly Lorentz gases consisting of an infinite number of convex (typically
circular) scatterers and the Ehrenfest gases consisting of an infinite number of polygons \cite{D,G98,MS}. Infinite periodic arrays in these models are equivalent to motion on a torus
and can be treated using similar techniques to those of equivalent finite billiards
\cite{DWY,GG}. However, very little has been proven about the more physically realistic
models with randomly placed non-overlapping obstacles \cite{T10}.  There is numerical
evidence for some statistical properties corresponding to a limiting Weiner (diffusion)
process in some Ehrenfest gases \cite{DC01,JBR}.

\begin{problem}
What are minimum dynamical properties required for a ``chaotic'' macroscopic limit, for
example recurrence, ergodicity, a ``normal'' diffusion coefficient, a limiting Weiner process?  
\end{problem}

While most of the above work pertains to closed systems (albeit on large or infinite domains),
it is worth pointing out that the ``escape rate formalism'' \cite{D,G98} relates escape rates
in large open chaotic systems to linear transport properties; note that the escape rate is also
related to other dynamical properties by Eq. (\ref{e:erf}). A final problem for this topic:

\begin{problem}
What experimental techniques might best probe the limits of statistical
mechanics arising from the previous two open problems?
\end{problem}

\paragraph{Quantum chaos}
Billiards correspond to the classical (short wavelength) limit of wave equations
for light, sound or quantum particles in a homogeneous cavity.  The classical dynamics
corresponds to the small wavelength, geometrical optics approximation. Semiclassical
theory uses properties of this classical dynamics, especially periodic orbits,
as the basis for a systematic treatment of the wave properties (eigenvalues and
eigenfunctions of the linear wave operator).  Comparisons are also made with the
predictions of random matrix theory, in which the spacing of eigenvalues of many quantum
systems follows universal laws based only on the chaoticity and symmetries of the
problem \cite{PDLMR}.%
\footnote{There are fascinating conjectures \cite{CFKRS} relating random matrix theory and the
Riemann zeta function of footnote \ref{f:RH}.}
Billiards are both necessary for calculating properties of wave systems and a testbed
for semiclassical and random matrix theories.  A useful survey of quantum chaos, including
open quantum systems, is given in \cite{N08}; note that semiclassical approaches are also
relevant to non-chaotic systems such as polygonal billiards \cite{HHM}.

An important recent development in open quantum systems has been the fractal Weyl conjecture
relating the number of resonances up to a particular energy to fractal dimensions of
classically trapped sets.  Following \cite{LSZ} we recall that a quantum billiard is
equivalent to the Helmholtz equation $(\nabla^2+k^2)\psi=0$ with a Dirichlet condition
$\psi=0$ at the boundary, the eigenvalues $k$ are labelled $k_n$ and are real.
In this case Weyl's law gives a detailed statement to the effect that the number of
states with $k_n<k$ is proportional to $Vk^d$ in the limit $k\to\infty$ where $d$ is
the dimension of the space; effectively this means that each quantum state occupies the
same classical phase space volume.

If the billiard is open, for example consisting of a finite region containing obstacles
that do not prevent the particle escaping to infinity, the Helmholtz equation has
resonance solutions where $k$ has negative imaginary part.  We have \cite{LSZ}

\begin{problem}
The fractal Weyl conjecture: The number of states with $Re(k_n)<k$ and $Im(k_n)>-C$
for some positive constant $C$ is of order $k^{d_H+1}$ where again $k\to\infty$ and $d_H$
is the partial Hausdorff dimension of the non-escaping set.
\end{problem}

While open billiards of the non-eclipsing type have been under investigation from the
beginning \cite{S90}, the most progress to date has been in simpler systems such as
quantum Baker maps \cite{KNNS,NPWCK}.  Recent work has also included smooth
Hamiltonian systems \cite{RPBF} and optical billiards \cite{WM}.

\paragraph{Experiments and further applications}
The case of microresonators where light is trapped by total internal reflection is of interest
for practical applications in laser design; the hole then corresponds to a condition on momentum
rather than position, and desired properties typically include large $Q$-factor (small imaginary
part of $k$) so that the pumping energy required for laser action is small, together with a
strongly directional wave function at infinity.  The earliest example of a circular cavity has
large $Q$-factor due to trapping of light by total internal reflection, but the symmetry precludes
a directional output, so a number of efforts have been made to modify the circular geometry, using insight from classical billiards \cite{ACWM,DMSW,POWR}.

Other experiments of relevance have included microwaves, sound, atoms and electrons in cavities at
scales ranging from microns to metres.  These are aimed at demonstrating the capabilities of new
experimental techniques, testing theoretical results from quantum chaos, and laying the groundwork
for specific applications.  Often modifications to the original billiard problem arise.  For example,
open semiconductor billiards can be constructed by confining a two dimensional electron gas (2DEG)
using electric fields.  In transport through such cavities, a weak applied magnetic field shifts
the quantum phases (hence conductance) while the classical orbits are effectively unchanged,
while a strong field also curves the classical orbits; also the walls are likely to be described
by a somewhat soft potential energy function.  For room acoustics, collisions involve a proportion
of the sound escaping, being absorbed, being randomly scattered, and being reflected using the
usual billiard law, depending on the materials at the relevant point on the boundary.  Relevant
references may be found in \cite{BD07,DG,H,WW}.

Open dynamical systems that are more general than billiards are important for escape and
transport problems in heat conduction \cite{GG}, chemical reactions \cite{EWW},
astronomy \cite{WBW} and nanotubes \cite{JBR}.  In these systems, open billiards provide a
useful starting point for an understanding of more general classes of open dynamical systems.

\begin{problem}
Generalisations: Billiards with more realistic physical effects, for example
soft wall potentials, external fields, dissipative and/or stochastic scattering (at
the boundary or in the interior) or time dependent boundaries.  
\end{problem}

The possibilities are endless; for this reason, it is increasingly important to
carefully justify the mathematical and/or physical interest of any new model.

\section{Discussion}

We conclude with a discussion of some problems that draw together many classes
of dynamics.
 
\begin{problem}
Exact expansions: For a given open (generalised) billiard problem, can the survival
probability $P(t)$ be expressed exactly, or at least as an expansion for large $t$
that goes beyond the leading term?
\end{problem}

The most likely candidates for this problem are those billiards that are best understood,
the integrable and dispersing cases.

One feature that is common to all classes of billiards, classical and quantum is
the role of periodic orbits in the open problem.  Periodic orbits denote an exact
recurrence to the initial state of a classical system, and so contribute directly
to the set of orbits that enter from a hole and return there after one period.
Periodic orbits may be of measure zero, but some positive measure neighbourhood
of a periodic orbit will be sufficiently close as to have the same property.
For (finite) billiards with no periodic orbits the Poincar\'e recurrence theorem
guarantees similar behaviour.

For an initial measure supported other than on the hole, periodic orbits are relevant
to the set of orbits which never escapes, and its neighbourhood often determines
the long time survival probability.  This is seen in many of the above sections,
for integrable, other polygonal, chaotic and intermittent billiards.  There is
substantial existing theory for calculating the escape rate $\gamma$ and
other long time statistical properties (averages etc.) for hyperbolic systems,
using either periodic orbits avoiding the hole \cite{L} or passing through the
hole \cite{AT}, and also for semiclassical treatments of quantum systems \cite{N08}.

\begin{problem}
Recurrence: Give a general description of how the escape problem is affected by
recurrences in the system (periodic orbits or more general), in the vicinity
of the hole and/or elsewhere, for systems with little or no hyperbolicity.
\end{problem}

Finally, a question posed in~\cite{BD05} but for which a systematic solution does not appear
to exist in the literature:

\begin{problem}
Inverse problems: What information can be extracted about the dynamics of a billiard
from escape measurements? Can you ``peep'' the shape of a drum? \cite{BD07,K66}.
\end{problem}

The open problems we have considered vary from the long standing and difficult
(problem \ref{o:tri})
to those arising very recently from active research that may as quickly solve them
(problem \ref{o:loc}).
Even the simplest of integrable systems, the circle, has unexpected complexity,
while a rigorous approach to hyperbolic billiards has a highly developed
and technical theory, and the study of polygons with irrational angles is in
its infancy.  It is likely that progress will be made almost immediately on many of
the problems in specific cases but that constructing reasonably complete solutions
in general dynamical contexts will be a active
area of research for many years.  Finally, open billiards are interesting from a
mathematical point of view and also required for solving practical problems; this
would appear to be a particularly fruitful field for collaborations between
mathematicians and physicists.

\section*{Acknowledgements}
The author owes much to patient collaborators and other colleagues including Eduardo Altmann,
Peter B\'alint, Leonid Bunimovich, Eddie Cohen, Orestis Georgiou, Alex Gorodnik, Jon Keating,
Edson Leonel, Jens Marklof and Martin Sieber for discussions on many topics related to this
article.

\end{document}